# General Effect Modelling (GEM)

## Part 1. Method description

Mosleth, E.F.[1] and Liland, K.H.[2]

[1]Nofma AS, Norwegian Institute of Food, Fisheries and Aquaculture Research, Osloveien 1, 1430 Ås,
Norway
[2]Faculty of Science and Technology, Norwegian University of Life Sciences, 1430 Ås,
Norway.

## Abstract

Modern analysis technologies output large amounts of multivariate data. The data may come from an experimental design or from other collected observations. We here present a flexible tool, called General Effect Modelling (GEM), for the analysis of any type of multivariate data influenced by one or more qualitative (categorical) or quantitative (continuous) input variables. The variables do not have to be design factors. They can also be observed values, for example, age, sex, or income, or they may represent subgroups of the samples discovered through data exploration. The first step in GEM separates the variation in the multivariate data into effect matrices associated with each of the influencing variables (and possibly interactions between these) by applying a general linear model. The residuals of the model are added to each of the effect matrices and the results are called Effect plus Residual (ER) values. The tables of ER values have the same dimensions as the original multivariate data. The second step of GEM is a multivariate or univariate exploration of the ER tables to learn more about the multivariate data in relation to each input variable. The multivariate exploration is simplified as it addresses one input variable at the time, or if preferred, a combination of the input variables. One example of the use of GEM is a study to identify molecular fingerprints associated with a disease that is not influenced by age where individuals at different ages with and without the disease are included in the experiment. Analysed by GEM, this situation can be described as an experiment with two input variables: *the targeted disease* and *the individual age*. Through GEM, the effect of age can be removed, thus focusing further analysis on the targeted disease without the influence of the confounding effect of age. ER values can also be the combined effect of several input variables, where influences of irrelevant variation, for example, running batch are removed. This publication has three parts where the first part, the present article, gives a description of the GEM methodology, and Part 2 is a consideration of multivariate data and the benefit of treating such data by multivariate analysis, with a focus on omics data, and Part 3 is a case study in Multiple Sclerosis (MS) where GEM is applied.

## Keywords

## Introduction

Multivariate analysis of data that originates from designed experiments with multiple factors is a challenge. Classical univariate analysis of designed experimental studies typically consists of ANalysis Of VAriance (ANOVA) for qualitative variables and linear regression models for quantitative variables. These univariate analyses can be extended to multivariate data by Multivariate ANOVA (MANOVA). MANOVA has, however, several limitations making it less suited for multicollinear response data [1]. On the other hand, many methods designed for multivariate multicorrelated data, are not well suited for experimental designed data with multiple design variables as the effect of the different design variables may be confused.

Several methods are developed to handle situations with multiple input design variables and multivariate response [2-7]. Some of these methods are based on ANOVA or regression on coded designs combined with multivariate analysis. By ANOVA, regression coefficients are estimated for each design variable reflecting the impact of each design variable on the observed data. The regression coefficients for an effect multiplied with the corresponding (coded) design variable result in an effect matrix, also known as a Least Squares estimate, of the design variable. Applied on all response variables, this results in a decomposition of the original data into one effect matrix for each design variable plus a residual matrix which contains the variation that is not explained by the design variables in the model. One approach for the analysis of data with multiple design variables and multivariate responses is to combine ANOVA with the explorative multivariate analysis Principal Component Analysis (PCA). One variant is ANOVA-PCA[6] and another is ASCA (Analysis of Variance Simultaneous Component Analysis) [8, 9]. Both methods first decompose the original data into effect matrices by ANOVA, but they differ in the way PCA is applied to the effects matrices. With ANOVA-PCA, the residual matrix is first added to the effect matrix and then the resulting array is decomposed by PCA, while with ASCA, PCA is applied to individual effect matrices only, and the differences between the replicates are utilized thereafter. The significance of factors can be estimated using permutation testing[10]. A newer approach is based on classical multivariate ANOVA to produce informative confidence ellipsoids in the score plots which also enables testing for significant differences between effect levels[4]. This gives validation of the design variables, and for balanced designs, this leads to exact significances, both for main effects and interactions. Validation of the individual response variables is, however, not directly available by these methods.

Alternative models exchange PCA with the supervised methods Partial Least Squares (PLS) [11], also called Projection to Latent Spaces (PLS), or Target Projection (TP) [12]. The latter uses a PLS model where a single latent variable (the target-projected component) is created from normalised predictions and their projection on the input matrix (effect matrix from ANOVA) [3, 7].

Whereas the above methods are linked to specific multivariate analyses, we recently presented a new data analysis method called Effect plus Residual (ER) modelling [2, 13], which allows any multivariate and univariate analysis to be applied. We hereby present a new method as an umbrella over different methods that analyse data with multiple input variables and multivariate response data, hereby called General Effect Modelling (GEM). Both quantitative and qualitative design variables are allowed in GEM. The variables do not have to be design factors. They can also be observed values, for example, age, sex, or income, or they may represent subgroups of the samples discovered through data exploration.

By GEM, the effects of each variable are first estimated for each response variable with the design variables as input, to simplify the subsequent statistical analysis. A General Linear Model (GLM) is used to estimate the effects and the residuals of the GLM model are added to each effect matrix to yield what is called Effect plus Residual (ER) values, similar to the use of ANOVA in ANOVA-PCA to obtain effect matrices with added residuals. Any univariate or multivariate model can then be applied to the ER values to explore the multivariate patterns related to one design variable at a time. This includes methods from the domains of Chemometrics, Data Mining, Machine Learning, and in general, the field of Artificial Intelligence. GEM can therefore be considered as an umbrella enabling a wider platform of methods developed for multivariate response data, encompassing ASCA, APCA, ANOVA-TP, and other tested and untested options. Leveraging a wider range of GEM possibilities, like mixed models, nested models, etc., and basing the calculations on regression, also ASCA+ [14] and LiMM-PCA [15] are within the umbrella.

## Theory

In the following, we will denote a response variable as $y_i$, $i = 1, \ldots, N$, where N is the number of responses. The collection of all responses, i.e., the observed data is denoted $Y$. A design variable is denoted as $x_d$, where $d$ is an integer corresponding to a single variable, e.g., $x_1$ for the first input variable, or a combination of integers corresponding to an interaction between variables, e.g., $x_{1:2:3}$ for the interaction between input variables 1, 2 and 3. For categorical variables, $x_d$ will be a matrix with sum-coding (explained below) with the number of columns one less than the number of categorical levels. Model errors for each response variable are denoted as $\epsilon_i$.

GEM consists of three main steps.

**Step 1** in GEM is the estimation of effects by linear modelling with the design variables as input and one response at a time as output in the model. This model is not used for statistical validation but for adjustment of the data to reflect the effects of only one design variable at a time, to simplify the subsequent statistical analysis. By the linear model, a regression coefficient, $\beta_{i,d}$, is estimated for each combination of input variable and response variable. In practice, $\beta_{i,d}$ may contain more than one value if a categorical variable with more than two levels is used. The collection of regression coefficients for all responses we denote $\boldsymbol{\beta}_d = [\beta_{1,d}, \ldots, \beta_{i,d}, \ldots, \beta_{N,d}]$. A linear model with two experimental factors, including the two main effects and their interaction effect (full factorial experiment), becomes:

$$y_i = x_1 \beta_{i,1} + x_2 \beta_{i,2} + x_{1:2} \beta_{i,1:2} + \epsilon_i \quad (1)$$

An illustration of a full factorial experiment with two design variables, each at two levels, and three biological replicates is given in Table 1. A corresponding example with two and three factor levels and two biological replicates is shown in the Appendix.

Table 1. Full factorial experimental design of two factors, each on two levels, and three biological replicates. The interaction effect is obtained by multiplying the two design variables, and it reflects different responses of one design variable depending on the level of the other design variable.

|  | $x_1$ | $x_2$ | $x_{1:2}$ | Replicate |
|---|---|---|---|---|
| sample 1 | -1 | -1 | **1** | 1 |
| sample 2 | -1 | -1 | **1** | 2 |
| sample 3 | -1 | -1 | **1** | 3 |
| sample 4 | -1 | **1** | -1 | 1 |
| sample 5 | -1 | **1** | -1 | 2 |
| sample 6 | -1 | **1** | -1 | 3 |
| sample 7 | **1** | -1 | **1** | 1 |
| sample 8 | **1** | -1 | **1** | 2 |
| sample 9 | **1** | -1 | **1** | 3 |
| sample 10 | **1** | **1** | -1 | 1 |
| sample 11 | **1** | **1** | -1 | 2 |
| sample 12 | **1** | **1** | -1 | 3 |

The regression coefficients, $\boldsymbol{\beta}_d$, are estimated for all responses of each design variable. The design variable multiplied with the corresponding regression coefficients gives the effects of each variable $\boldsymbol{E}_d = \boldsymbol{x}_d \boldsymbol{\beta}_d$. For the model in Equation (1), the combined effects of all input variables can be written as follows:

$$\boldsymbol{Y} = \boldsymbol{x}_1 \boldsymbol{\beta}_1 + \boldsymbol{x}_2 \boldsymbol{\beta}_2 + \boldsymbol{x}_{1:2} \boldsymbol{\beta}_{1:2} + \boldsymbol{R} = \boldsymbol{E}_1 + \boldsymbol{E}_2 + \boldsymbol{E}_{1:2} + \boldsymbol{R}, \tag{2}$$

where $\boldsymbol{R}$ is the matrix of residuals, not captured by the GLM effects.

From this, we combine the effect matrices with the residual matrix to form Effect plus Residual (ER) matrices. These contain the patterns relating to one design element and the residuals of the full model (Equation (2)(1)). Using the same example model we obtain the following three ER matrices:

$$\begin{aligned} \boldsymbol{Y} &= \boldsymbol{E}_1 + \boldsymbol{E}_2 + \boldsymbol{E}_{1:2} + \boldsymbol{R} \\ \boldsymbol{ER}_1 &= \boldsymbol{E}_1 \phantom{+ \boldsymbol{E}_2 + \boldsymbol{E}_{1:2}} + \boldsymbol{R} \\ \boldsymbol{ER}_2 &= \phantom{\boldsymbol{E}_1 +} \boldsymbol{E}_2 \phantom{+ \boldsymbol{E}_{1:2}} + \boldsymbol{R} \\ \boldsymbol{ER}_{1:2} &= \phantom{\boldsymbol{E}_1 + \boldsymbol{E}_2 +} \boldsymbol{E}_{1:2} + \boldsymbol{R} \end{aligned} \tag{3}$$

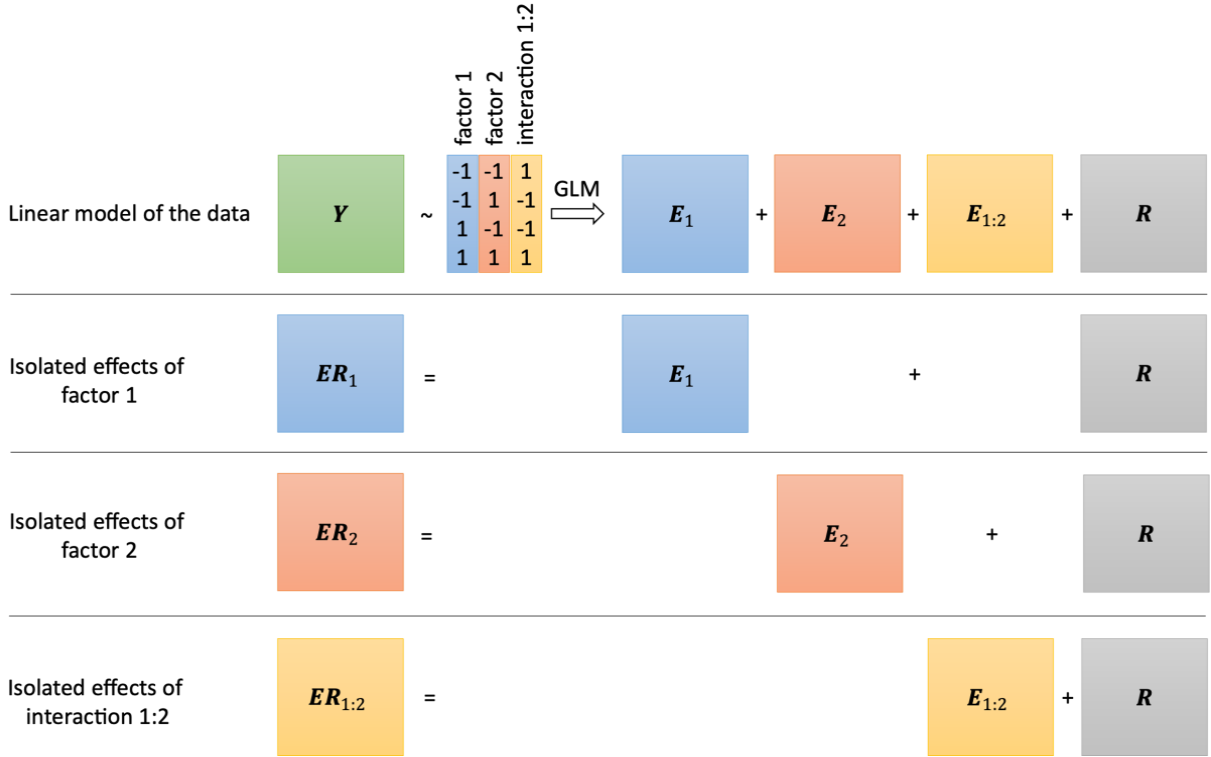

*Figure 1 Graphical representation of the first step of GEM.*

**Step 2** in GEM is to perform univariate or multivariate exploration or analyses of the ER matrices. Typical explorations include visualisations of single ER variables to gain insights about distributions, uncertainties, etc. or combinations of variables to look for interactions or multivariate patterns. To obtain an overview of all ER variables simultaneously, PCA is an excellent tool which decomposes the information into variable contributions and sample patterns which can be visualised as loading ($\boldsymbol{P}$) plots and score ($\boldsymbol{T}$) plots. More details are given in the Component Methods section. Scores and loadings are in general obtained as:

$$\boldsymbol{T}_d, \boldsymbol{P}_d := PCA(\boldsymbol{ER}_d) \qquad (4)$$

More targeted analyses can be performed using PLS [2, 11, 16] or Elastic Net [17] analyses with an ER matrix as input and the corresponding design variable as output. PLS scores, loading weights and loadings are in general obtained as:

$$\boldsymbol{T}_d, \boldsymbol{W}_d, \boldsymbol{P}_d := PLS(\boldsymbol{x}_d \sim \boldsymbol{ER}_d). \qquad (5)$$

More details about PLS are given in the Component Methods section. Subsequent analyses like variable selection can then be performed to gain insight into which response variables influence the design variables. For PLS, computation of variable importance or variable selection is most commonly a post-processing step, e.g., using Jackknifing[18] or other variable selection method [19, 20], while for Elastic Net the use of L1 shrinkage gives an inherent variable selection. And as mentioned in the Introduction, any analyses of $\boldsymbol{ER}_d$ and $\boldsymbol{x}_d$ that can shed light on the problem at hand is an option for further analysis, i.e.:

$$\begin{aligned}\ldots &:= f(\boldsymbol{ER}_d) \\ \ldots &:= g(\boldsymbol{x}_d \sim \boldsymbol{ER}_d)\end{aligned} \qquad (6)$$

## Component-based methods

PCA and PLS are two different component-based methods, that find useful linear combinations of the original variables in lower dimensional spaces. They are bilinear in that both samples and variables are decomposed simultaneously into scores and loadings, respectively. Both provide informative views into high-dimensional data by projecting the data onto low-dimensional components that capture relevant phenomena and variation. To illustrate their differences, a toy example is shown in Figure 2 where variables x1 and x2 are related to a response y. The colours of the samples indicate the value of y. PCA identifies the directions that span most of the variation in the combination of x1 and x2, illustrated by blue lines. The first component is the long, primary axis, while the second component is the short, secondary axis orthogonal to the first one. In contrast, PLS identifies the directions that covary the most with the response(s). In the figure, we can observe the effect in the rotated red axes that align more with the colour change in the samples. The plot legend also shows the proportion of explained variance of the data (X) for the first component for PCA and PLS and the proportion of explained variance of the response (y) when used for regression. For PCA, using the components for regression is called Principal Component Regression, and PLS regression is often abbreviated PLSR. As expected, PCA is best at recovering the data, while PLS is best at predicting the response.

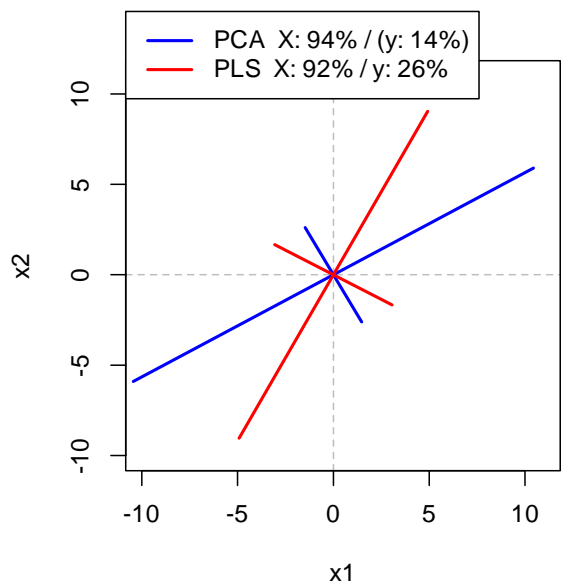

*Figure 2 Comparison of Principal Component Analysis and Partial Least Squares in a simple two-dimensional case.*

## Regularisation methods

Regularisation can be applied to a regression or classification method to penalise large regression coefficients, thereby stabilising the solution and possibly crystalising out important variables. If the focus is only on stabilising, L2 shrinkage, i.e., limiting the sum of squared regression coefficients, as applied in Ridge Regression and Tikhonov Regression, can be implemented very efficiently [21]. If the goal is to reduce complexity by zeroing contributions from many variables, combining L2 shrinkage with L1 shrinkage, i.e., limiting the sum of absolute regression coefficients, results in the well-known Elastic net.

## Example

The following example applies the R package gemR to perform GEM including multivariate analysis by PLS and Elastic Net.

```r
# The example assumes that the following objects are available in
# R's Global Environment:
# proteins - numeric matrix of proteins
# ms       - character vector of MS ("no"/"yes")
# group    - numeric vector of groups discovered by cluster analysis
# sex      - character vector of sex ("M"/"F")
# age      - numeric vector of patient ages

# Step 0 - Prepare data
# Organise data as data.frame.
    MS.data <- data.frame(proteins = I(proteins),
                ms = factor(ms),
                group = factor(group),
                sex = factor(sex),
                age = age)

# Step 1 - GLM:
    MS.gem <- GEM(proteins ~ ms + group + sex + age, data = MS.data)

# Matrices of effects and ER values can be extracted for custom analyses.
# Here shown for the 'ms' effect
    E.ms  <- MS.gem$effects$ms
    ER.ms <- MS.gem$ER.values$ms

# Step 2 - PLS
    ncomp <- 10
# PLS analysis of the 'ms' effect with Jackknifing
    ms.pls <- pls(MS.gem, 'ms', ncomp,
                jackknife = TRUE)

# Element extraction
# Cross-validated classifications of 'ms' (# correct)
    colSums(ms.pls$classes == as.numeric(MS.data$ms))

# Scores, loadings and coefficients for 'ms' effect
    ms.scores   <- scores(ms.pls)
```

```r
    ms.loadings <- loadings(ms.pls)
    ms.coef     <- coef(ms.pls)

# Parsimonious estimate of #components
    ms.ncomp    <- 2
# Jackknife P-value estimates, significant and subset
    ms.jack     <- ms.pls$jack[,1,ms.ncomp]
    ms.signif   <- ms.jack < 0.05
    ms.subset   <- which(ms.signif)

# Example plots
    oldpar <- par(mfrow = c(2,2), mar = c(4,4,2,1), mgp = c(2.5, 1, 0))
    scoreplot(ms.pls, main = "Scores (ms)",
        panel.first = abline(h=0, v=0, col="gray"),
        col = MS.data$ms, pch = as.numeric(MS.data$group), cex = 0.7)
    loadingplot(ms.pls, scatter = TRUE, main = "Loadings (ms)",
        col = c("gray", "black")[ms.signif+1], pch = c(1,15)[ms.signif+1],
        panel.first = abline(h=0, v=0, col="gray"), cex = 0.7)
    corrplot(ms.pls, main = "Correlation loadings",
        col = c("gray", "black")[ms.signif+1], pch = c(1,15)[ms.signif+1],
        cex = 0.3)
    plot(ms.pls, ylim = c(0,100), cex = 0.7)
    par(oldpar)
```

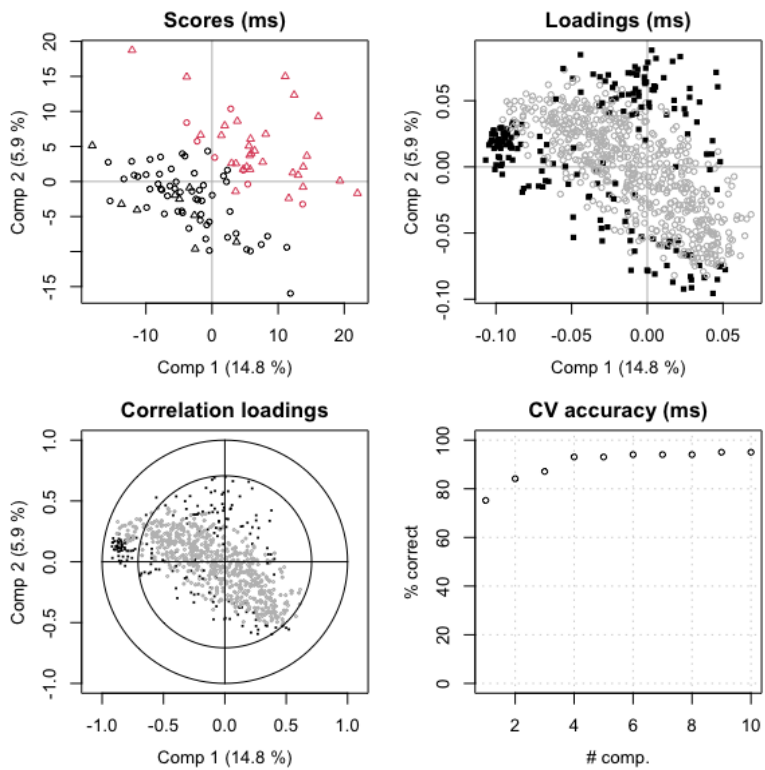
*Figure 3 Example plots for PLS analysis in GEM using 'ms' as response.*

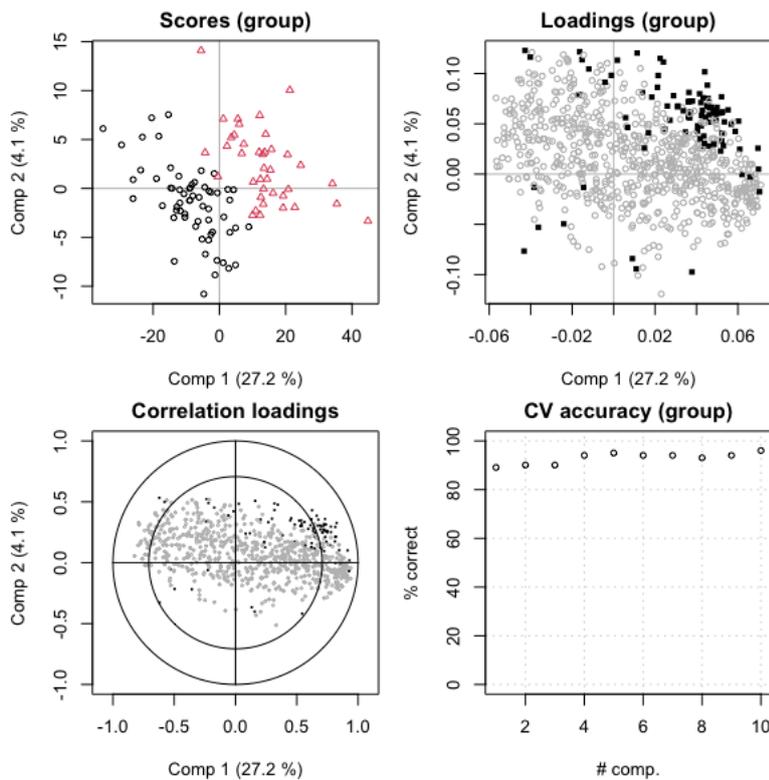
*Figure 4 Example plots for PLS analysis in GEM using 'group' as response (code not shown).*

```
# PLS analysis of the 'ms' effect with Shaving (repeated sMC selection)
    ms.pls.sh       <- pls(MS.gem, 'ms', ncomp,
```

```r
              shave = TRUE)

# Simplest model that retains maximum performance and corresponding variables
    min.red       <- ms.pls.sh$shave$min.red
    optimal.subset <- ms.pls.sh$shave$variables[[min.red+1]]

# Find the error level of classifying all samples as the majority class/level
    maxClass    <- function(x){x <- table(x); 1-max(x)/sum(x)}
    bestProp    <- maxClass(ms.pls.sh$gem$data[[ms.pls.sh$effect]])

# Plot shaving results
    plot(ms.pls.sh, ylim = c(0, bestProp), ylab = "% error", main = "Shaving")
    abline(h = bestProp, lty = 2); grid()
    legend(26,0.35, legend = "Majority class error", lty = 2)
```

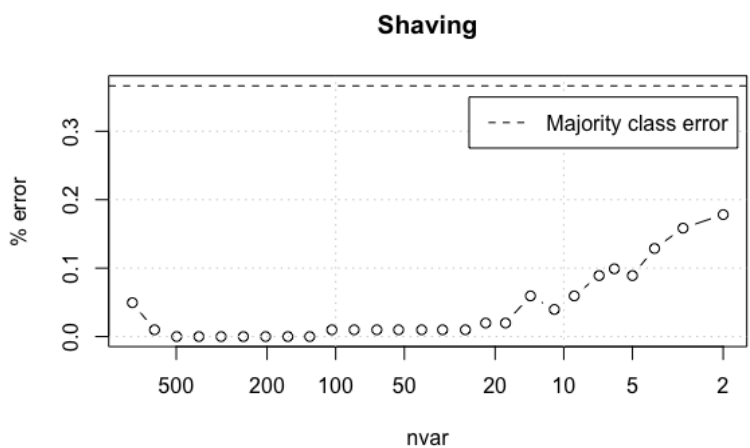

*Figure 5 Effect of shaving (default parameters) one classification error as a function of the number of remaining variables.*

```r
# Step 2 - Elastic Net
# Elastic Net analysis of the 'group' effect
    MS.en.group <- elastic(MS.gem, 'group', validation = "LOO",
                alpha = 0.5, family = "binomial")

# Element extraction
# Coefficients for 'group' effect
    MS.coef.group <- coef(MS.en.group)
# Non-zero coefficients at optimal complexity
    nonZeroID  <- which(MS.coef.group[-1,1] != 0)
    nonZeroLab <- names(nonZeroID)
```

```
# Example plots
oldpar <- par(mfrow = c(3,1), mar = c(4,4,2,1), mgp = c(2.5, 1, 0))
plot(MS.en.group$glmnet, xvar = "lambda")
plot(MS.en.group)
par(oldpar)
```

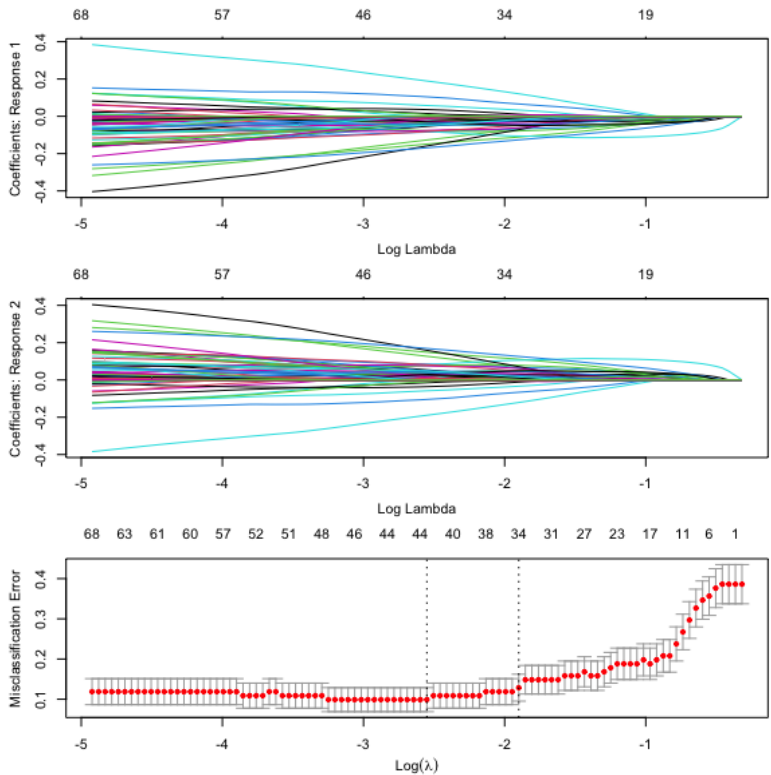

*Figure 6 Example plots for Elastic Net in GEM.*

## Author contributions

Ellen Mosleth and Kristian Hovde Liland contributed equally to the development of GEM, Kristian Hovde Liland programmed the accompanying R package, and Ellen F Mosleth performed the analyses in R.

## Competing interests

The authors declare no competing interests.

# Appendix/Supplementary

Table A1: Full factorial experimental design of two factors having two and three levels, respectively, and two biological replicates.

|            | $x_1$ | $x_2$ |    | $x_{1:2}$ |    | Replicate |
|------------|-------|-------|----|-----------|----|-----------|
| sample 1   | -1    | -1    | -1 | 1         | 1  | 1         |
| sample 2   | -1    | -1    | -1 | 1         | 1  | 2         |
| sample 3   | -1    | 0     | 1  | 0         | -1 | 1         |
| sample 4   | -1    | 0     | 1  | 0         | -1 | 2         |
| sample 5   | -1    | 1     | 0  | -1        | 0  | 1         |
| sample 6   | -1    | 1     | 0  | -1        | 0  | 2         |
| sample 7   | 1     | -1    | -1 | -1        | -1 | 1         |
| sample 8   | 1     | -1    | -1 | -1        | -1 | 2         |
| sample 9   | 1     | 0     | 1  | 0         | 1  | 1         |
| sample 10  | 1     | 0     | 1  | 0         | 1  | 2         |
| sample 11  | 1     | 1     | 0  | 1         | 0  | 1         |
| sample 12  | 1     | 1     | 0  | 1         | 0  | 2         |